\documentclass[aps,prl,twocolumn,floatfix,superscriptaddress,showpacs]{revtex4}

\usepackage{epsfig}
\usepackage{amsmath}
\usepackage{bm}

\begin{document}

\title{Elliptic flow of thermal photons in relativistic nuclear collisions}
\author{Rupa Chatterjee}
\affiliation{Variable Energy Cyclotron 
Centre, 1/AF Bidhan Nagar, Kolkata 700 064, India}            
\author{Evan S.~Frodermann}
\affiliation{Physics Department, The Ohio State University, 
             Columbus, OH 43210, USA} 
\author{Ulrich Heinz}
\affiliation{Physics Department, The Ohio State University, 
             Columbus, OH 43210, USA} 
\author{Dinesh K.~Srivastava}
\affiliation{Variable Energy Cyclotron 
Centre, 1/AF Bidhan Nagar, Kolkata 700 064, India}            
\date{\today}

\begin{abstract}
We predict the transverse momentum ($p_T$) dependence of elliptic flow 
of thermal photons for Au+Au collisions at the Relativistic Heavy Ion 
Collider. We model the system hydrodynamically, assuming formation 
of a thermalized quark-gluon plasma at some early time, followed by 
cooling through expansion, hadronization and decoupling. Photons are 
emitted throughout the expansion history. Contrary to hadron elliptic 
flow, which hydrodynamics predicts to increase monotonically with 
$p_T$, the elliptic flow of thermal photons is predicted to first 
rise and then fall again as $p_T$ increases. Photon elliptic flow at 
high $p_T$ is shown to reflect the quark momentum anisotropy at early 
times when it is small, whereas at low $p_T$ it is controlled by the 
much larger pion momentum anisotropy during the late hadronic emission 
stage. An interesting structure is predicted at intermediate 
$p_T{\,\sim\,}0.4$\,GeV/$c$ where photon elliptic flow reflects 
the momenta and the (compared to pions) reduced $v_2$ of heavy vector 
mesons in the late hadronic phase.
\end{abstract}

\maketitle
Experiments performed at the Relativistic Heavy-Ion Collider (RHIC)
are providing evidence for the production of quark-gluon plasma (QGP) 
in nuclear collisions at ultra-relativistic energies. Key findings 
include strong anisotropic flow of all hadronic species 
\cite{v2_exp,v2_part,v2_theo} and a suppression of high-$p_T$ hadrons 
due to parton energy loss in the dense medium \cite{jetq_theo,jetq_exp}. 
Signatures of direct photon emission \cite{guy,simon,bms_phot,fms_phot},
indicative of a hot early state, have also started emerging \cite{phenix1}. 
The emphasis of the next generation of experiments will thus necessarily 
shift to a more precise determination of the properties of both the QGP and 
the subsequent hot hadronic matter. Hydrodynamic flow, and in particular
anisotropic flow in non-central collisions, provides strong evidence
for the existence of a hot and dense initial state with thermal pressure 
\cite{v2_theo}. Elliptic flow is generated very early, via the 
transformation of the initial spatial eccentricity of the nuclear overlap
region into momentum anisotropies, through the action of azimuthally 
anisotropic pressure gradients. With the passage of time, the pressure
gradients equalize, and the growth of elliptic flow shuts itself off
\cite{sorge}.

Due to their strong interactions, hadrons decouple from the system 
late, typically when the temperature has dropped to values around 
100\,MeV \cite{v2_theo,QGP3}. In hydrodynamic simulations, hadrons 
with large transverse momenta are thus emitted from those fluid 
elements which have the largest radial flow. This is supported 
by the hydrodynamically predicted $p_T$-dependence of the elliptic
flow coefficient $v_2$ and of the Hanbury-Brown Twiss radii~\cite{QGP3}.
Photons, on the other hand, are emitted at every stage of the 
collision, from the pre-equilibrium stage~\cite{bms_phot}, the quark-gluon
fluid, and the late hadronic matter. The thermal emission of photons 
from the QGP and hadronic phases is obtained \cite{KKMR} by integrating 
the thermal emission rate (which is strongly biased towards higher 
temperatures) over the space-time history of the system. As a result, 
high-$p_T$ photons arise mostly from the hot early stage, where 
hydrodynamic flow is weak but the spatial eccentricity of the source 
is large, whereas hadrons are emitted when the temperature is low, 
the flow is strong and anisotropic, but the spatial eccentricity of 
the fireball has mostly disappeared. The elliptic flow of photons, 
especially at $p_T{\,\agt\,}1{\,-\,}2$\,GeV/$c$, is therefore expected 
to provide a glimpse of the early part of the expansion history when
the fireball is in the QGP phase, complementary to the elliptic flow 
of hadrons. 

This last argument can be made independent of the validity of 
hydrodynamics. For two of the leading photon production processes 
(quark-gluon Compton scattering and quark-antiquark annihilation for high 
energy photons in the QGP, $\pi\pi{\,\to\,}\rho\gamma$ for low energy 
photons in the hadronic phase) it is known that, for photon energies well 
above the rest masses of the emitting particle, the photon production cross 
section peaks very strongly for momenta close to that of the emitting 
particle \cite{cyw}. So, even without local thermal equilibrium, 
one expects the photon elliptic flow to track the momentum anisotropy 
of the photon-emitting particles at similar momenta. 

For Au+Au collisions at RHIC it is known \cite{v2_part,RANP04} that the 
hydrodynamic behaviour of elliptic flow begins to break down for mesons 
(baryons) above $p_T{\,\simeq\,}1.5$\,GeV/$c$ ($2.3$\,GeV/$c$): instead 
of continuing to rise with $p_T$, as predicted by hydrodynamics 
\cite{v2_theo}, the elliptic flow saturates. On the other hand, the 
systematics of hadron production with $p_T{\,\agt\,}2$\,GeV/$c$ can 
be well described by quark coalescence \cite{coal}. The observations 
then translate into a quark elliptic flow near hadronization that breaks 
away from hydrodynamics above $p_T{\,\simeq\,}0.75$\,GeV/$c$ \cite{RANP04}. 
Thus, even though ideal fluid dynamics excellently describes the bulk of 
particle production at RHIC, viscous corrections become significant 
at transverse momenta above 1\,GeV/$c$ for quarks and gluons and above 
$2-3$\,GeV/$c$ for hadrons. We expect these non-equilibrium features
to be reflected by the photons, and the hydrodynamic prediction for 
photon elliptic flow presented here to be only an upper limit once 
$p_T$ exceeds 1 GeV/$c$. Nevertheless, the qualitative features 
pointed out below are generic and expected to be robust against 
non-equilibrium corrections.  
  
The photon momentum spectrum can be written as 
\begin{equation}
\label{eq1}
  E\, dN_\gamma/d^3p = \int\bigl[(...)\exp(-p{\cdot}u(x)/T(x))\bigr]\,d^4x\,.
\end{equation}
The square bracket indicates the thermal emission rate from the QGP or 
hadronic matter. 
$p^\mu{\,=\,}(p_T \cosh Y, p_T\cos\phi,p_T\sin\phi,p_T \sinh Y)$ denotes 
the 4-momentum of the photon, while the 4-velocity of the flow field is  
$u^\mu = \gamma_T\bigl(\cosh \eta,v_x(x,y),v_y(x,y),$ $\sinh\eta\bigr)$ 
(with $\gamma_T = (1{-}v_T^2)^{-1/2}$, $v_T^2{\,=\,}v_x^2{+}v_y^2$),
assuming boost-invariant longitudinal 
expansion. We use coordinates $\tau,\,x,\,y,\,\eta$, with 
volume element $d^4x{\,=\,}\tau\, d\tau \, dx \, dy\, d\eta$, where 
$\tau{\,=\,}(t^2{-}z^2)^{1/2}$ is the longitudinal proper time and
$\eta{\,=\,}\tanh^{-1}(z/t)$ is the space-time rapidity. The photon 
momentum is parametrized by its rapidity $Y$, transverse momentum 
$p_T{\,=\,}(p_x^2+p_y^2)^{1/2}$, and azimuthal emission angle 
$\phi$. The photon energy in the local fluid rest frame, which enters 
as $p{\cdot}u/T$ into the Boltzmann and several other factors in the 
thermal emission rate, is given by
\begin{equation}
\label{eq2}
 \frac{p \cdot u}{T} = \frac{\gamma_T p_T}{T} 
 \bigl[\cosh (Y{-}\eta)- v_T\cos(\phi{-}\phi_v)\bigr]\,,
\end{equation}
where $\phi_v{\,=\,}\tan^{-1}(v_y/v_x)$ is the azimuthal angle of the
tranverse flow vector. Eq.~(\ref{eq2}) shows that the azimuthal 
anisotropy ($\phi$-dependence) of the photon spectrum, conventionally 
characterized by its Fourier coefficients $v_n$ (where for equal nuclei 
only even $n$ contribute at $Y{\,=\,}0$),
\begin{equation}
\frac{dN(b)}{d^2p_T \, dY}=\frac{dN(b)}{2 \pi p_T \, dp_T \, dY}
\left[1+2 v_2(p_T,b) \cos(2 \phi)+\dots\right],
\nonumber
\end{equation}
is controlled by an interplay between collective flow anisotropy
and geometric deformation of the temperature field $T(x,y,\tau)$: 
It obviously vanishes in the absence of radial flow, $v_T{\,=\,}0$, 
but for nonzero $v_T$ it can arise from an anistropic flow field 
or from an azimuthally deformed temperature field. Only if the 
transverse flow points radially ($\phi_v{\,=\,}\phi_r{\,=\,}\tan^{-1}(y/x)$)
{\em and} both the flow and temperature fields are azimuthally symmetric 
($v_T(x,y){\,=\,}v_T(r)$ and $T(x,y){\,=\,}T(r)$ with $r^2{\,=\,}x^2{+}y^2$) 
can the $\phi$-dependence in Eq.~(\ref{eq2}) be made to disappear, by
rotating in (\ref{eq1}) $\bm{r}{\,\equiv\,}(x,y){\,=\,}(r,\phi_r)$ 
such that $\phi_r{\,=\,}\phi$.

We employ the boost invariant hydrodynamic code AZHYDRO \cite{AZHYDRO}
which has been used extensively to explore hadron production at 
RHIC \cite{QGP3}. We use standard \cite{QGP3,AZHYDRO} initial 
conditions for Au+Au collisions at $\sqrt{s}{\,=\,}200\,A$\,GeV,
but extrapolated from the usual initial time $\tau_0{\,=\,}0.6$\,fm/$c$
to a 3 times smaller value of $\tau_0{\,=\,}0.2$\,fm/$c$,
assuming 1-dimensional boost-invariant expansion between these times.
We do so in order to account for at least a fraction of the 
pre-equilibrium photon production at very early times \cite{BMS04};
its contribution to the photon spectrum is important at large $p_T$,
and will suppress $v_2^\mathrm{photon}$ there because very little 
transverse flow develops before 0.6\,fm/$c$. So our initial maximum 
entropy density in the center of the fireball for $b{\,=\,}0$ collisions
is $s_0{\,=\,}3\times117\,\mathrm{fm}^{-3}{\,=\,}351\,\mathrm{fm}^{-3}$,
corresponding to a peak initial temperature of $T_0{\,=\,}520$\,MeV.
The initial transverse entropy density profile is computed
from the standard Glauber model parametrization \cite{QGP3}.

%
\begin{figure}
\centerline{\epsfig{file=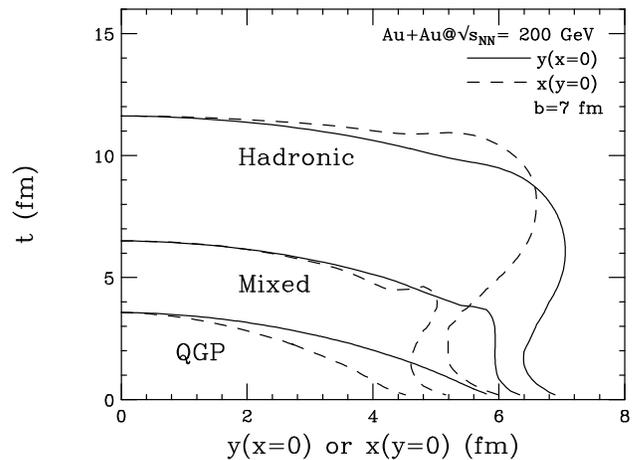,width=0.95\linewidth}}
\caption{Constant energy density contours for
 $\epsilon=\epsilon_\mathrm{q}$, $\epsilon_\mathrm{h}$, and 
 $\epsilon_\mathrm{f}$, along $y(x{=}0)$ and $x(y{=}0)$,
 for Au+Au at $b{\,=\,}7$\,fm.
\vspace*{-3mm}
}
\label{fig1}
\end{figure}
%

We assume creation of a thermally and chemically equi\-li\-brated plasma
at $\tau_0$, and use the complete leading-order rate for QGP photon 
production from Ref.~\cite{guy} and the latest results for photon 
radiation by a hot hadron gas from Ref.~\cite{simon}. These rates 
provide good descriptions of single photon data at SPS energies 
and recently at RHIC \cite{fms_phot,dks_int,EP05}. We use the equation 
of state EOS\,Q \cite{AZHYDRO} which matches a non-interac\-ting QGP to 
a chemically equilibrated hadron gas at $T_c{\,=\,}164$\,MeV, 
with energy densities $\epsilon_\mathrm{q}(T_c){\,=\,}1.6$ GeV/fm$^3$ and 
$\epsilon_\mathrm{h}(T_c){\,=\,}0.45$\,GeV/fm$^3$ in the subphases. 
Hadron freeze-out is assumed to happen at 
$\epsilon_\mathrm{f}{\,=\,}0.075$\,GeV/fm$^3$ \cite{QGP3}.

\begin{figure}
\centerline{\epsfig{file=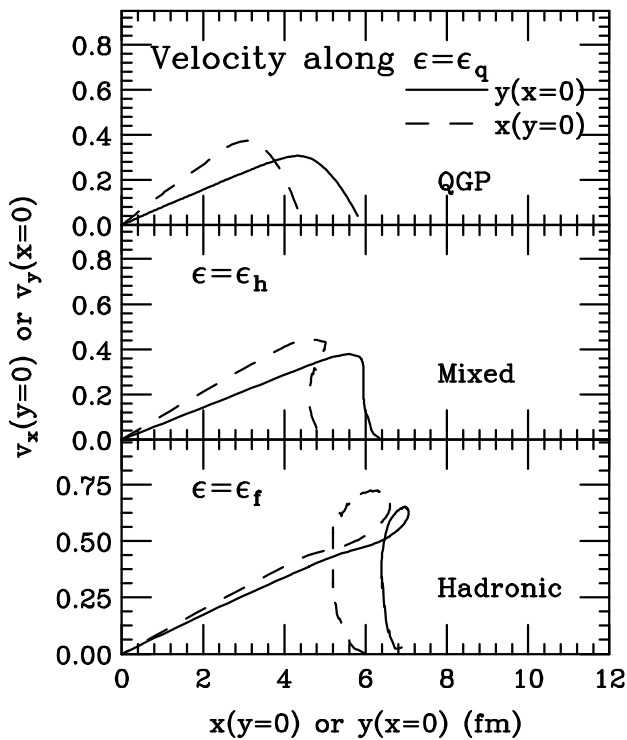,width=0.95\linewidth}}
\caption{Flow velocity along the constant energy density contours for
 $\epsilon{\,=\,}\epsilon_\mathrm{q}$ (QGP phase, upper panel),
 $\epsilon_\mathrm{h}$ (mixed phase, middle panel), and
 $\epsilon_\mathrm{f}$ (hadronic phase, lower panel) for 
 $y(x{=}0)$ and $x(y{=}0)$, for 200\,$A$\,GeV Au+Au at $b{\,=\,}7$\,fm.
\vspace*{-2mm}
}
\label{fig2}
\end{figure}

Figure~\ref{fig1} shows the changing spatial anisotropy of the fireball,
for Au+Au collisions at a typical impact parameter of $b{\,=\,}7$\,fm,
by plotting cuts through constant energy density surfaces along the 
$x$ and $y$-axes (i.e. in the reaction plane and perpendicular to it)
for the three values $\epsilon_\mathrm{q}$, $\epsilon_\mathrm{h}$, and 
$\epsilon_\mathrm{f}$. After about 9\,fm/$c$ the initial out-of-plane 
deformation changes sign (dashed and solid lines cross), as a result of 
faster expansion into the reaction plane. Figure~\ref{fig2} shows
the flow velocities along the same cuts, illustrating the developing
anisotropy of the transverse flow field. Through most of the fireball
interior the velocity along the $x$-axis is larger and rises more 
rapidly with increasing radius than along the $y$-axis. At hadronic 
freeze-out, $\epsilon_\mathrm{f}$, the in-plane and out-of-plane 
velocity gradients have almost (but not fully!) equalized, and 
larger in-plane flow velocities exist mostly near the fireball surface. 
Figure~\ref{fig2} complements similar plots in Refs.~\cite{v2_theo,QGP3} 
which show velocity profiles either at constant time or only along the 
freeze-out surface at $b{\,=\,}0$.  

%
%

\begin{figure}
\centerline{\epsfig{file=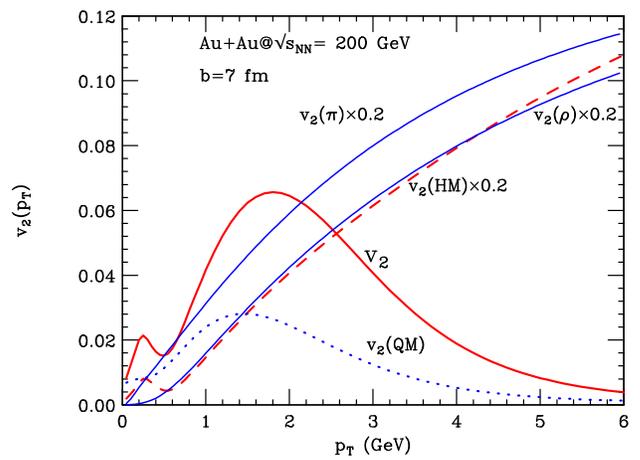,width=0.95\linewidth}}
\caption{(Color online) 
$v_2(p_T)$ for thermal photons from off-central 200\,$A$\,GeV Au+Au 
collisions at $b{\,=\,}7$\,fm.
Quark and hadronic matter contributions are also shown separately,
and the elliptic flow of $\pi$ and $\rho$ mesons is shown for comparison.
\vspace*{-2mm}
}
\label{fig3}
\end{figure}

In Fig.~\ref{fig3} we show the thermal photon elliptic flow for 
Au+Au collisions at $b{\,=\,}7$\,fm. We compare the overall $v_2(p_T)$ 
with the individual elliptic flow coefficients associated with quark 
matter ($v_2(\mathrm{QM})$) and hadronic matter ($v_2(\mathrm{HM})$) 
contributions to the thermal $\gamma$ spectrum. Comparison of 
$v_2(\mathrm{HM})$ with the elliptic flow of thermal $\pi$ and $\rho$ 
mesons, $v_2(\pi)$ and $v_2(\rho)$, shows that at low (high) $p_T$ 
hadronic photon flow tracks the elliptic flow of $\pi$ ($\rho$) mesons. 
This can be understood from the single-photon spectrum (see, e.g., Fig.~17 
in \cite{QGP3_EM}): at low $p_T$ hadronic photon emission is dominated by 
the processes $\rho{\,\to\,}\pi\pi\gamma$ and $\pi\pi{\,\to\,}\rho\gamma$ 
whereas at higher $p_T$ {\em collision-induced conversion of vector mesons into 
photons} (such as $\pi\rho{\,\to\,}\pi\gamma$) takes over. In the
present situation, with radial flow effects on the spectra computed
from hydrodynamics, this transition happens around 
$p_T{\,\sim\,}0.4$\,GeV/$c$. At $p_T{\,\gtrsim\,}0.4$\,GeV/$c$ hadronic
photon elliptic flow thus tracks the $v_2$ of $\rho$ and other vector 
mesons. Since the heavier vector mesons carry less elliptic flow than 
the lighter pions \cite{v2_theo}, this transition manifests itself in a 
reduction of hadronic photon $v_2$ around $p_T{\,\sim\,}0.4$\,GeV/$c$
(see dashed red line in Fig.~\ref{fig3}). Due to the dominance of 
hadronic processes in the total photon yield in this $p_T$ region, this
structure survives in the total photon elliptic flow (solid red line). 

%
%

The elliptic flow of the QGP photons, $v_2(\mathrm{QM})$, is small at 
low $p_T$, peaks around $p_T{\,\sim\,}1.5$\,GeV/$c$, and decreases again 
for large $p_T$. Its smallness at high $p_T$ and increase towards lower 
$p_T$ reflects the absence of transverse flow during the earliest, 
hottest stage of the QGP and its gradual buildup during the following 
cooler stages. That growth is cut off by the generic decrease of $v_2$ 
as $p_T{\,\to\,}0$ \cite{D95}. Its nonzero limit at $p_T{\,=\,}0$ 
(first observed in \cite{HW02} for gluons) can be traced to the 
singularities of the Bose distribution and emission rates at 
$p_T{\,=\,}0$. Since the QGP photon rates \cite{guy} are unreliable 
for $p_T{\,<\,}0.2$\,GeV/$c$, the $v_2(\mathrm{HM})$ curve in 
Fig.~\ref{fig3} should not be trusted in that domain. Fortunately, 
there the total $v_2$ (solid red line) is entirely dominated by the 
hadronic photon contribution.  

Although the hydrodynamically predicted elliptic flow of hadronic 
photons is almost everywhere much larger than that of the QGP photons,
the hadronic contribution to the photon spectrum is increasingly 
suppressed (by more than an order of magnitude below the QGP 
contribution) once $p_T$ exceeds $1.5{\,-\,}2$\,GeV/$c$. Hence the 
overall photonic $v_2$, while larger than the pure QGP contribution, 
decreases for large $p_T$, too, approaching the $v_2$ of the QGP 
photons. At high $p_T$ the photon elliptic flow thus opens a window onto
the dynamics of the QGP, in spite of the larger elliptic flow of 
the hadronic photons.

\begin{figure}
\centerline{\epsfig{file=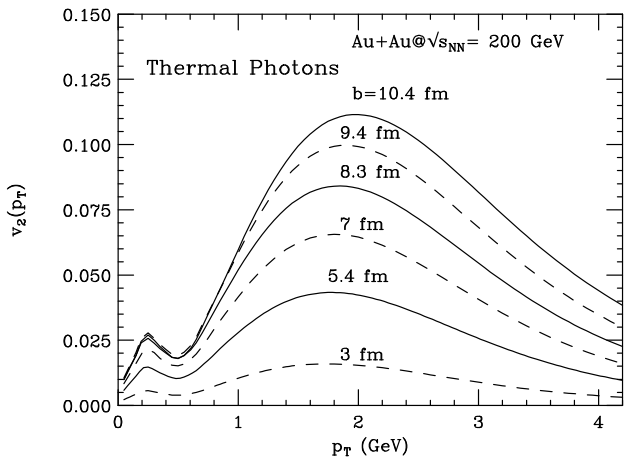,width=0.95\linewidth}}
\caption{Impact parameter dependence of the azimuthal anisotropy of thermal
photons.
\vspace*{-2mm}
}
\label{fig4}
\end{figure}

In Figure~\ref{fig4} we present the impact parameter dependence of the
elliptic flow of thermal photons. The impact parameters are chosen to
roughly correspond to collision centralities of $0{\,-\,}10$\%, 
$10{\,-\,}20$\%, $20{\,-\,}30$\%, $30{\,-\,}40$\%, $40{\,-\,}50$\%, 
and $50{\,-\,}$60\% of the total inelastic nuclear cross section. As 
the impact parameter increases, the relative contribution of hadronic 
photons increases, too. Since its elliptic flow is larger than that 
of the QGP photons, the $b$-dependence of thermal photon 
$v_2$ is predicted to be stronger than that of hadron $v_2$. This can 
be used as an additional tool to isolate the QGP contribution to 
thermal photon emission, especially if methods can be developed 
\cite{phenix2} to subtract at high $p_T$ the hadronic contribution 
$v_2(\mathrm{HM})$ from the thermal photon $v_2$, by exploiting its 
similarity to the elliptic flow of thermal $\rho$ mesons. 

%
%
 
However, a quantitative interpretation of photon elliptic flow in
the context of extracting dynamical information on the QGP evolution 
must also account for other contributions to high-$p_T$ photon 
production. It has recently been argued that jets passing through 
the QGP contribute significantly to direct photon production at 
large $p_T$ \cite{fms_phot} and may even lead to negative photon
elliptic flow in this region \cite{gale}. It will be interesting to 
see how much of this effect survives once the so far neglected 
anisotropic collective flow of the medium is included.

In conclusion, we have presented a first calculation of elliptic flow of
thermal photons from ultrarelativistic heavy-ion collisions. The 
differential photon elliptic flow $v_2(p_T)$ exhibits a rich structure,
driven by the evolution of the system and the 
competing rates of photon emission from the quark and hadronic matter
stages. Its impact parameter dependence differs from that of pion elliptic
flow since the relative contributions of hadronic and quark matter 
emissions change with collision centrality. At high $p_T$, where 
ideal fluid dynamics is known to overpredict $v_2$, the elliptic flow 
of thermal photons from the hadronic matter tracks the flow of $\rho$
mesons which can be measured independently. If this can be used to subtract 
the hadronic contribution to photon elliptic flow at high $p_T$, a new 
window onto the early buildup of collective flow in the QGP will be opened. 
 
This work was supported by the U.S. Department of Energy under 
contract no. DE-FG02-01ER41190.


\end{document}